\def\be{\begin{equation}}
\def\ee{\end{equation}}
\def\ba{\begin{array}}
\def\ea{\end{array}}

\def\L{\Lambda}
\def\l{\lambda}

\documentclass[prl,amsfonts,showpacs,twocolumn,pra]{revtex4}
\usepackage{graphicx}
\usepackage{caption2}
\usepackage{float}
\usepackage{amsmath}
\usepackage{amssymb}
\usepackage[OT2,OT1]{fontenc}
\usepackage{hyperref}
\usepackage[hyphenbreaks]{breakurl}
\usepackage{mathrsfs,color,float,indentfirst,textcomp}
\usepackage{latexsym,lscape,enumerate}
\def\qed{\leavevmode\unskip\penalty9999\hbox{}\nobreak\hfill
     \quad\hbox{\leavevmode  \hbox to.77778em{%
               \hfil\vrule   \vbox to.675em%
               {\hrule width.3em\vfil\hrule}\vrule\hfil}}
     \par\vskip 0pt}

\begin{document}
\setcounter{secnumdepth}{3} 
\renewcommand\thesection{\Roman{section}}
\renewcommand\thesubsection{\Alph{subsection}}
\title{Uncertainties of Genuinely Incompatible Triple Measurement Based on Statistical Distance}
\author{Hui-Hui Qin$^{1}$}
\author{Ting-Gui Zhang$^{2}$}
\author{Leonardo Jost$^{3}$}
\author{Chang-Pu Sun$^{1,4}$}
\author{Xianqing Li-Jost$^{2,5}$}
\author{Shao-Ming Fei$^{5,6}$}

\affiliation{$^1$ Beijing Computational Science Research Center, Beijing 100193, P. R. China\\
$^2$ School of Mathematics and Statistics, Hainan Normal University, Haikou 571158, China\\
$^3$ Universit\"{a}t Regensburg, Universit\"{a}tsstrasse 31, Regensburg 93053, Germany\\
$^4$ Graduate School of China Academy of Engineering Physics,
Beijing 100193, P. R. China\\
$^5$ Max-Planck-Institute for Mathematics in the Sciences, Leipzig 04103, Germany\\
$^6$ School of Mathematical Sciences, Capital Normal University,
Beijing 100048, China}

\begin{abstract}
We investigate the measurement uncertainties of a triple of positive operator-valued measures (POVMs) based on statistical distance, and formulate state-independent tight uncertainty inequalities satisfied by the three measurements in terms of triple-wise joint measurability. Particularly, uncertainty inequalities for three unbiased qubit measurements are presented with analytical lower bounds which relates to the necessary and sufficient condition of the triple-wise joint measurability of the given triple. We show that the measurement uncertainties for a triple measurement are essentially different from the ones obtained by pair wise measurement uncertainties by comparing the lower bounds of different measurement uncertainties.
\end{abstract}


\maketitle
\section{Introduction}
The uncertainty principle is arguably one of the most famous features of quantum mechanics \cite{Heisenberg}, which limits the accuracy of measuring some properties of a quantum system.
The well-known Heisenberg-Robertson uncertainty relation says that \cite{Robertson29}, for any observables $A$ and $B$, $\Delta A\Delta B\geq\frac{1}{2}|\langle[A,B]\rangle|$,
where $\Delta\Omega=\sqrt{\langle \Omega^{2}\rangle-\langle \Omega\rangle^{2}}$ is the standard deviation for observable $\Omega$, $\langle\cdot\rangle$ denotes the expectation of an operator with respect to a given state $\rho$, and $[A,B]=AB-BA$. This state-dependent inequality implies the impossibility of simultaneously determining the definite values of non-commuting observables. Such uncertainty relations based on product form or summation form of deviation have been generalized and studied \cite{Weigert,chenprl,qinhh,Dammeier,Buschbook,
Kech2,Maccone,Pati,Chensum}. The entropic uncertainty relations \cite{Dammeier,Maassen,Berta,Coles,Fan,xiaoyl,Wehener,
Riccardi} and measurement
probability based universal uncertainty relations \cite{y13,y14,y15,y16,Barchielli,Barchielli2,Kech}, with or without quantum memory, have been extensively investigated. Besides, uncertainty relations based on measurement noise and disturbance have been also derived and experimentally verified \cite{Ozawa03,ozawa2004uncertainty,Buscemi,Sulyok}.

Since the influence of the measurement on quantum systems is not always the reason for uncertainty \cite{ff}, there are uncertainty relations, of which the uncertainties are described by approximation error for probabilities of joint measurements \cite{Barchielli,Barchielli2,Busch07,Busch13PRL,Busch14PRA,
Busch14RMP,Ma}. In \cite{Barchielli,Barchielli2} the approximation error for probabilities is quantified by the sum of relative entropies, while in \cite{Busch07,Busch13PRL,Busch14PRA,
Busch14RMP,Ma} the corresponding approximation error for probabilities is quantified by $L_1$-distances. In addition, in \cite{Barchielli,Barchielli2} multi spin-$1/2$ components measurement uncertainty relations have been studied. In
\cite{Busch07,Busch13PRL,Busch14PRA,Busch14RMP,Ma} two measurement uncertainty relations have been investigated. Since a triple measurement uncertainty relation deduced from a two observable uncertainty relation \cite{chenprl} is usually not tight, triple measurement uncertainty relations are essentially different from the ones obtained by pair wise measurement uncertainties: there exist genuine incompatible triple measurements such that they are pair-wise jointly measurable, just like the case of genuine tripartite entanglement or genuine non-local correlations.

In this paper, based on statistical distance we formulate state-independent tight uncertainty relations satisfied by three measurements in terms of their triple-wise joint measurability. By approximating a given triple of unbiased qubit measurements to all possible triple measurements that are triple-wise jointly measurable, we show that the approximation error is lower bounded by a quantity which relates to the necessary and sufficient condition of the triple-wise joint measurability of the given triple. We also compare the different uncertainty relations which are obtained by approximation of triple-wise jointly measurable measurements and pair-wise jointly measurable measurements, respectively. Examples are given to illustrate the merit our the uncertainty relation.

\section{Triple measurement uncertainty relation}
\begin{figure}
\centering
\includegraphics[width=0.5\textwidth]{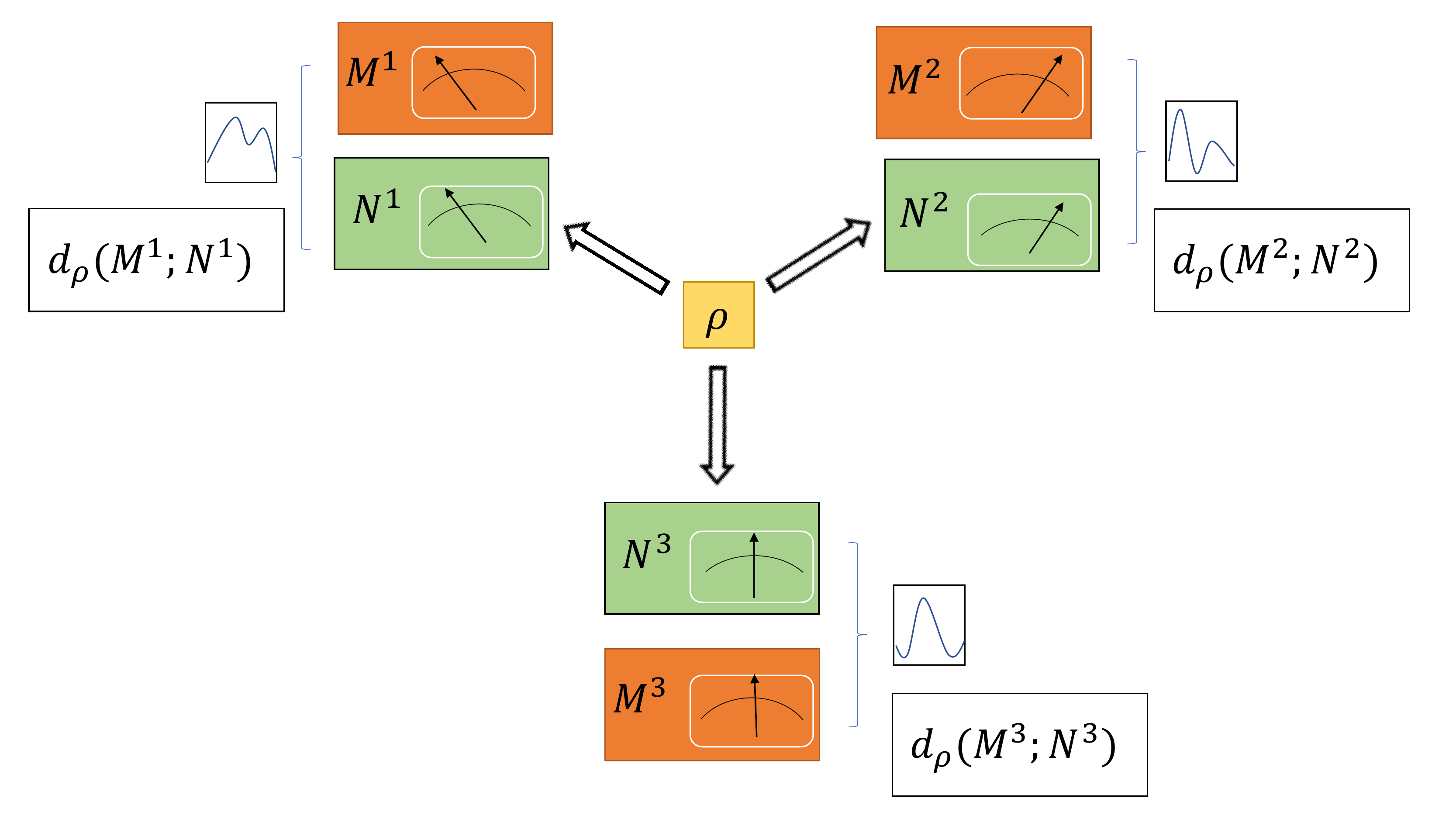}
\caption{(Color online) The approximation of $\{M^{i}\}^3_{i=1}$ to triple-wise
jointly measurable measurements $\{N^{i}\}^3_{i=1}$.}\label{Fig.2}
\end{figure}

Consider three positive operator-valued measures $\{M^{i}\}^{3}_{i=1}$, given by the
semi-positive measurement operators $\{M^{i}_{k}|\,M^{i}_{k}\geq 0,~ \sum_{k}M^{i}_{k}=\mathbb{I}\}$, $i=1,2,3$, where $\mathbb{I}$ stands for the identity operator.
Let $\{N^{i}_{k}|N^{i}_{k}\geq 0,~ \sum_{k}N^{i}_{k}=\mathbb{I}\}$, $i=1,2,3$, be another set of three positive operator-valued measures which are triple-wise jointly measurable.
For an arbitrary given state $\rho$, the measurement probabilities with respect
to $M^{i}_k$ ($N^{i}_k$) are given by $p^{M^i}_k=Tr(\rho M^i_k)$
($p^{N^i}_k=Tr(\rho N^i_k)$).
The approximation error between measurements $M^{i}$ and $N^{i}$ is given by $d_{\rho}(M^{i};N^{i}):=\sum_{k}|p^{M^{i}}_{k}
-p^{N^{i}}_{k}|$.
By maximizing $d_{\rho}$ over all $\rho$, we obtain state-independent approximation error, which is the worst case on all states, between the triple measurements $\{M^{1},M^{2},M^{3}\}$ and the triple-wise jointly measurable measurements $\{N^{1},N^{2},N^{3}\}$, i.e.,
\be\label{delta}
\Delta(M^1,M^2,M^3;N^1,N^2,N^3):=
\max_{\rho}\sum^{3}_{i=1}d_{\rho}(M^{i};N^{i}).
\ee
Let $\Delta_{lb}(M^1,M^2,M^3)$ denote the minimal value of $\Delta(M^1,M^2,M^3;N^1,N^2,N^3)$
over all possible triple-wise jointly measurable triples $N^1$, $N^2$ and $N^3$. Then the
quantity $\Delta_{lb}(M^1,M^2,M^3)$ quantifies the degree of incompatibility of the triple measurements $\{M^i\}^3_{i=1}$, see FIG. \ref{Fig.2}. It is apparent that $\Delta_{lb}(M^1,M^2,M^3)=0$ if and only if $M^1$, $M^2$ and $M^3$ are triple-wise jointly measurable.

Consider now three unbiased qubit measurements $\{M^{i}\}^{3}_{i=1}$ described by positive operator-valued measures
$$
M^{i}_{+}=\frac{I+\vec{m}_i\cdot\vec{\sigma}}{2},~~
M^{i}_-=\frac{I-\vec{m}_i\cdot\vec{\sigma}}{2}, ~~~{i=1,2,3},
$$
where the three dimensional vectors $\vec{m}_i$ satisfy $|\vec{m}_i|\leq 1$, $I$ is the $2\times 2$ identity matrix, and $\vec{\sigma}$ is the vector with the Pauli matrix $\sigma_i$ as the i-th entry. Let $\rho$ be a qubit state with Bloch vector representation, $\rho=(I+\vec{r}\cdot\vec{\sigma})/{2}$ ( $|\vec{r}|\leq 1$).
Maximizing $\sum^{3}_{i=1}d_{\rho}(M^{i};N^{i})$ over all $\rho$, we obtain
\be
\Delta(M^{1},M^{2},M^{3};N^{1},N^{2},N^{3})
=2\max_{\vec{r}}\sum^{3}_{i=1}|\vec{r}\cdot(\vec{m}_{i}-\vec{n}_{i})|.
\ee
For simplicity, in the following we denote $\vec{m}_{123}=\vec{m}_{1}+\vec{m}_{2}+\vec{m}_{3}$,
$\vec{m}_{ij}=\vec{m}_{i}+\vec{m}_{j}$, $\vec{n}_{123}=\vec{n}_{1}+\vec{n}_{2}+\vec{n}_{3}$ and
$\vec{n}_{ij}=\vec{n}_{i}+\vec{n}_{j}$.
It has been demonstrated in \cite{Yuarxiv} that three unbiased qubit measurements $\{N^i_{\pm}=(I\pm\vec{n}_i\cdot\vec{\sigma})/{2}\}^3_{i=1}$ are triple-wise jointly measurable if and only if
\be\label{tj}
\sum^4_{k=1}|\vec{p}_k-\vec{p}_F|\leq 4,
\ee
where
$\vec{q}_1=\vec{n}_{123}$, $\vec{q}_2=\vec{n}_1-\vec{n}_{23}$,
$\vec{q}_3=\vec{n}_2-\vec{n}_{13}$, $\vec{q}_4=\vec{n}_3-\vec{n}_{12}$ and $\vec{q}_F$ is the Fermat-Torricelli point of $\{\vec{q}_k\}^4_{k=1}$ \cite{BoltyanskiSpringer}. Minimizing $\Delta(M^{1},M^{2},M^{3};N^{1},N^{2},N^{3})$ under all triple-wise jointly measurable measurements $\{N^{i}\}^3_{i=1}$ satisfying (\ref{tj}), we have the following theorem,

{\bf Theorem} The approximation error of three unbiased qubit measurements $\{M^{i}\}^{3}_{i=1}$ to triple-wise jointly measurable unbiased qubit measurements $\{N^{i}\}^{3}_{i=1}$ satisfies the following inequality,
\be\label{jmerror}
\Delta(M^{1},M^{2},M^{3};N^{1},N^{2},N^{3})
\geq\frac{1}{2}(\sum^{4}_{k=1}|\vec{p}_{F}-
\vec{p}_{k}|-4),
\ee
where $\vec{p}_1=\vec{m}_{123}$, $\vec{p}_2=\vec{m}_1-\vec{m}_{23}$,
$\vec{p}_3=\vec{m}_2-\vec{m}_{13}$, $\vec{p}_4=\vec{m}_3-\vec{m}_{12}$ and $\vec{p}_F$ is the Fermat-Torricelli point of $\{\vec{p}_k\}^4_{k=1}$

{\sf [Proof]} By direct calculation we have the state-dependent approximation error,
\begin{widetext}
\be\label{Statedependent}
\begin{aligned}
&\sum^{3}_{i=1}d_{\rho}(M^{i};N^{i})
=2\sum^{3}_{i=1}|\vec{r}\cdot(\vec{m}_{i}-\vec{n}_{i})|\\
&\quad
 =2\cdot\left\{
 \begin{aligned}
  \displaystyle
  &\begin{aligned}
    \displaystyle
    &|\vec{r}\cdot(\vec{m}_{123}-\vec{n}_{123})|
    \leq|\vec{m}_{123}-\vec{n}_{123}|:=|\vec{g}_1|,\\
    \displaystyle
    &\qquad\mathrm{if}\quad [\vec{r}\cdot(\vec{m}_{1}-\vec{n}_{1})][\vec{r}
    \cdot(\vec{m}_{2}-\vec{n}_{2})]
    \geq0\wedge
   [\vec{r}\cdot(\vec{m}_{12}-\vec{n}_{12})]
   [\vec{r}\cdot(\vec{m}_{3}-\vec{n}_{3})]\geq0;
  \end{aligned}\\[2mm]
  \displaystyle
  &\begin{aligned}
    \displaystyle
    &|\vec{r}\cdot(\vec{m}_{1-23}-\vec{n}_{1-23})|
      \leq|\vec{m}_{1-23}-\vec{n}_{1-23}|:=|\vec{g}_{2}|,\\
    \displaystyle
    &\qquad\mathrm{if}\quad
    [\vec{r}\cdot(\vec{m}_{1}-\vec{n}_{1})]
    [\vec{r}\cdot(\vec{m}_{2}-\vec{n}_{2})]\leq0
    \wedge
   [\vec{r}\cdot(\vec{m}_{1-2}-\vec{n}_{1-2})]
   [\vec{r}\cdot(\vec{m}_{3}-\vec{n}_{3})]\leq0;
  \end{aligned}\\[2mm]
  \displaystyle
  &\begin{aligned}
     \displaystyle
     &|\vec{r}\cdot(\vec{m}_{2-13}-\vec{n}_{2-13})|
     \leq|\vec{m}_{2-13}-\vec{n}_{2-13}|:=|\vec{g}_{3}|,\\
     \displaystyle
     &\qquad\mathrm{if}\quad[\vec{r}\cdot(\vec{m}_{1}-
     \vec{n}_{1})][\vec{r}\cdot(\vec{m}_{2}-\vec{n}_{2})]
     \leq0\wedge
    [\vec{r}\cdot(\vec{m}_{1-2}-\vec{n}_{1-2})]\vec{r}
    \cdot(\vec{m}_{3}-\vec{n}_{3})]\geq0;
   \end{aligned}\\[2mm]
   \displaystyle
   &\begin{aligned}
    \displaystyle
    &|\vec{r}\cdot(\vec{m}_{3-12}-\vec{n}_{3-12})|
    \leq|\vec{m}_{3-12}-\vec{n}_{3-12}|:=|\vec{g}_{4}|,\\
    \displaystyle
    &\qquad\mathrm{if}\quad[\vec{r}\cdot
    (\vec{m}_{1}-\vec{n}_{1})][\vec{r}\cdot
    (\vec{m}_{2}-\vec{n}_{2})]
    \geq0\wedge
    [\vec{r}\cdot(\vec{m}_{1-2}-\vec{n}_{1-2})]
    [\vec{r}\cdot(\vec{m}_{3}-\vec{n}_{3})]\leq0.
   \end{aligned}
 \end{aligned}
 \right.
 \end{aligned}
\ee
\end{widetext}
We show that $\mathcal{G}:=2\max_i|\vec{g}_{i}|$, $i=1,2,3,4$, in (\ref{Statedependent}) can be reached.
Let $\rho_{0}$, with $\vec{r}=\vec{r}_{0}$, be the optimal state maximizing $\sum^{3}_{i=1}d_{\rho}(M^{i};N^{i})$.
Without loss of generality, we assume $\mathcal{G}=|\vec{g}_{1}|>0$. Set $\vec{r}_{0}={\vec{g}_{1}}/{|\vec{g}_{1}|}$, we have
\be\label{condition1.1}
 \begin{aligned}
 &[\vec{r}_{0}\cdot(\vec{n}_{1}-\vec{m}_{1})][\vec{r}_{0}\cdot(\vec{n}_{2}-\vec{m}_{2})]\\
 &=\frac{1}{|\vec{g}_{1}|^{2}}\Big[|\vec{n}_{1}-\vec{m}_{1}|^{2}+(\vec{n}_{23}-\vec{m}_{23})
 \cdot(\vec{n}_{1}-\vec{m}_{1})\Big]\\
 &\qquad\quad\cdot\Big[|\vec{n}_{2}-\vec{m}_{2}|^{2}
 +(\vec{n}_{13}-\vec{m}_{13})\cdot(\vec{n}_{2}-\vec{m}_{2})\Big]\\
 &\geq 0,
 \end{aligned}
\ee
where the inequality holds as $(\vec{n}_{23}-\vec{m}_{23})\cdot(\vec{n}_{1}-\vec{m}_{1})\geq0$ and
$(\vec{n}_{13}-\vec{m}_{13})\cdot(\vec{n}_{2}-\vec{m}_{2})\geq0$, since $|\vec{g}_{1}|\geq|\vec{g}_{2}|$ and $|\vec{g}_{1}|\geq|\vec{g}_{3}|$.

Similarly from $|\vec{g}_{1}|\geq|\vec{g}_{4}|$, $(\vec{n}_{12}-\vec{m}_{12})\cdot(\vec{n}_{3}-\vec{m}_{3})\geq0$, we have
 \be\label{condition1.2}
 \begin{aligned}
 &[\vec{r}\cdot(\vec{n}_{12}-\vec{m}_{12})][\vec{r}\cdot(\vec{n}_{3}-\vec{m}_{3})]\\
 &=\frac{1}{|g_{1}|^{2}}\Big[|\vec{n}_{3}-\vec{m}_{3}|^{2}+
 (\vec{n}_{3}-\vec{m}_{3})\cdot(\vec{n}_{12}-\vec{m}_{12})\Big]\\
 &\qquad\quad\cdot\Big[|\vec{n}_{12}-\vec{m}_{12}|^2+(\vec{n}_{12}-\vec{m}_{12})\cdot(\vec{n}_{3}-\vec{m}_{3})\Big]\\
 &\geq0.
 \end{aligned}
 \ee
(\ref{condition1.1}) and (\ref{condition1.2}) are just the first constraints in (\ref{Statedependent}).
Therefore, all together we have
 \be
 \begin{aligned}
\Delta&(M^1,M^2,M^3;N^1,N^2,N^3)\\
&=2\max\{|\vec{g}_{1}|,|\vec{g}_{2}|,
|\vec{g}_{3}|,|\vec{g}_{4}|\}:=2\mathcal{G}.
 \end{aligned}
 \ee
Noting that $\vec{g}_{i}=\vec{p}_{i}-\vec{q}_{i}$ and $\sum^4_{k=1}|\vec{q}_F-\vec{q}_k|\leq4$, we have
\be\label{ur3}
\begin{aligned}
\Delta&(M^1,M^2,M^3;N^1,N^2,N^3)=2\mathcal{G}\\
&\geq \frac{1}{2}\sum^4_{k=1}|\vec{p}_{k}-\vec{q}_{k}|
=\frac{1}{2}\sum^{4}_{k=1}|\vec{p}_{k}-\vec{q}_{F}+\vec{q}_{F}-\vec{q}_{k}|\\
&\geq\frac{1}{2}\sum^{4}_{k=1}[|\vec{p}_{k}-\vec{q}_{F}|-|\vec{q}_{F}-\vec{q}_{k}|]
\geq\frac{1}{2}[\sum^{4}_{k=1}|\vec{p}_{k}-\vec{p}_{F}|-4],
\end{aligned}
\ee
where the second inequality is due to triangle inequality, the third one comes from the definition of the Fermat-Torricelli
point of $\{\vec{p}_{k}\}^4_{k=1}$ and the constraint of the triple wise joint measurability for $\{N^i\}^3_{i=1}$. \qed

Apparently, if the lower bound of (\ref{jmerror}) is zero, then $M^{1},M^{2},M^{3}$ are triple-wise jointly measurable. From the definition of $\Delta_{lb}(M^{1},M^{2},M^{3})$ we then have $\Delta_{lb}(M^{1},M^{2},M^{3})=0
=\frac{1}{2}(\sum^{4}_{k=1}|\vec{p}_{F}-
\vec{p}_{k}|-4)$. It means that the inequality (\ref{jmerror}) is tight in the sense that the minimal value of $\Delta(M^{1},M^{2},M^{3};N^{1},N^{2},N^{3})$ is achieved by the lower bound. In this case the degree of the incompatibility of the given triple measurement is $0$. In the following we call a triple measurement $\{M^{1},M^{2},M^{3}\}$ a genuine incompatible triple measurement if the lower bound of (\ref{jmerror}) is strictly greater than zero.

\begin{figure}[H]
\centering
\includegraphics[width=1.0\columnwidth]{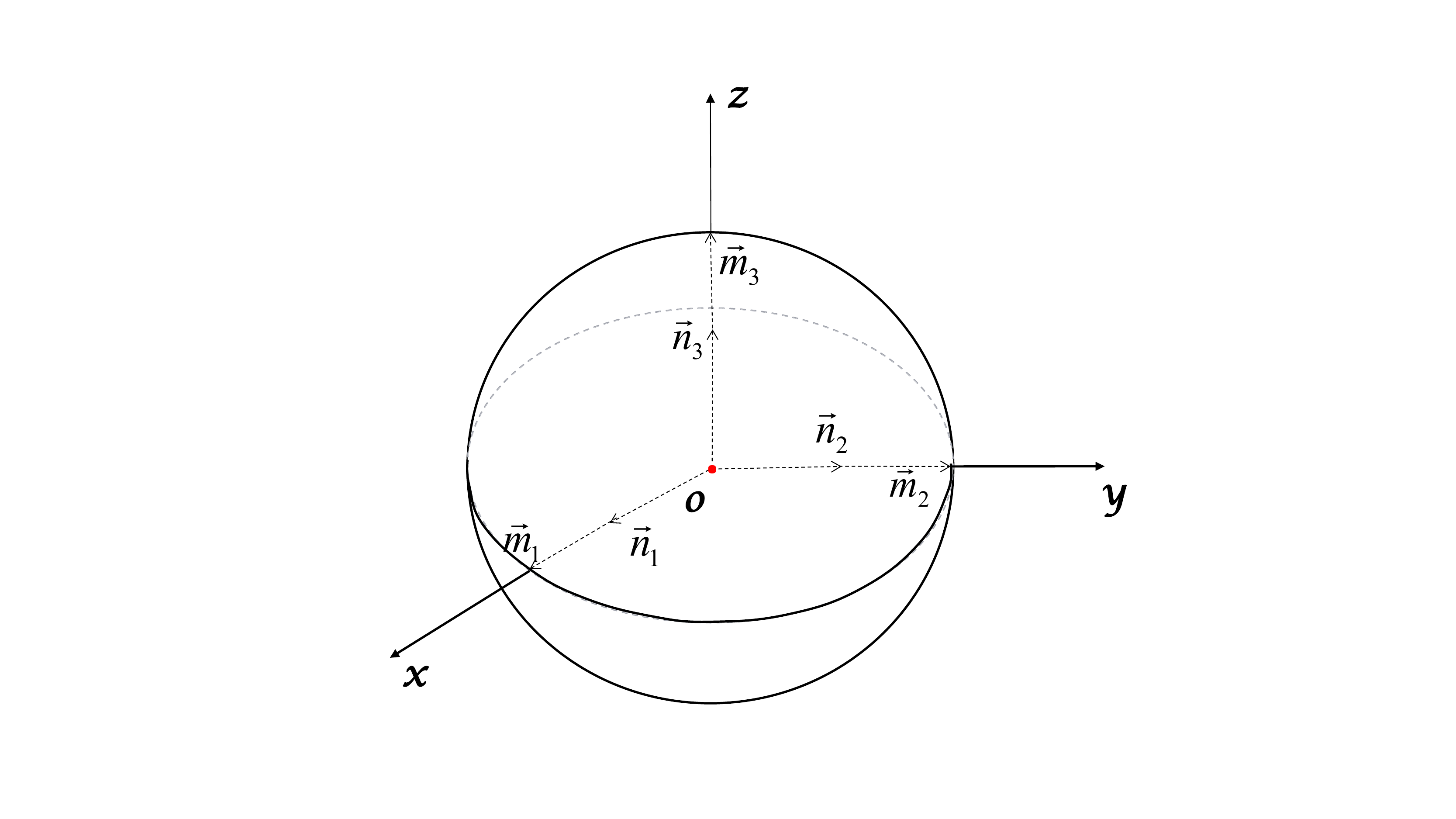}
\caption{(Color online) An optimal approximation of $\{M^{i}=\sigma_i\}^3_{i=1}$ by triple-wise jointly measurable $\{N^{i}\}^3_{i=1}$ given by $\vec{n}_i=\frac{1}{\sqrt{3}}\vec{m}_i$.}\label{Fig.4}
\end{figure}

Let us consider three sharp unbiased qubit measurements associated with the Pauli operators $\sigma_{i}$, $i=1,2,3$. Set
$\vec{m}_{1}=(1,0,0)$, $\vec{m}_{2}=(0,1,0)$ and $\vec{m}_{3}=(0,0,1)$. Then the three positive operator-valued measures $M^{1}$, $M^{2}$ and $M^{3}$ are just the projective measurements with respect to the eigenvectors of the three Pauli matrices, respectively. We have $\vec{p}_1=(1,1,1)$, $\vec{p}_2=(1,-1,-1)$, $\vec{p}_3=(-1,1,-1)$ and $\vec{p}_4=(-1,-1,1)$, which constitute a regular tetrahedron. And the Fermat-Torricelli point is exactly the origin, $\vec{p}_{F}=0$.
One can verify that the optimal approximation of triple-wise jointly measurable $\{N^i\}^{3}_{i=1}$ is given by $\vec{n}_i=\frac{1}{\sqrt{3}}\vec{m}_i$, as shown in FIG. \ref{Fig.4}.
The minimal value of
$\Delta(M^{1},M^{2},M^{3};N^{1},N^{2},N^{3})$ is actually the lower bound of (\ref{jmerror}), i.e.,
\be
\begin{aligned}
\Delta_{lb}(M^{1},M^{2},M^{3})
=\frac{1}{2}(\sum^{4}_{k=1}|\vec{p}_{k}|-4)=
2\sqrt{3}-2.
\end{aligned}
\ee
Therefore, the uncertainty inequality (\ref{jmerror})
is tight not only in trivial case but also tight in this case. Thus the triple measurement $\{M^1,M^2,M^3\}$ is genuine incompatible triple measurement and it's degree of incompatibility is $2\sqrt{3}-2$.

\section{Uncertainty: triple-wise versus pair-wise joint measurement approximation}
We next investigate the difference between  measurement uncertainty relations which are obtained by minimizing
$\Delta(M^1,M^2,M^3;N^1,N^2,N^3)$ over pair-wise and triple-wise jointly measurable
measurements, respectively. In \cite{Busch13PRL,Busch14PRA,Busch14RMP} this kind of Heisenberg's error-disturbance relation for a pair of measurements has been studied. For a given pair of measurements $M^{1}$ and $M^{2}$, their approximation to a pair of jointly measurable measurements $N^{1}$ and $N^{2}$, $\Delta(M^1,M^2;N^1,N^2):=\max_{\rho}\sum^{2}_{i=1}
d_{\rho}(M^{i};N^{i})$, satisfies the following relation \cite{Ma},
\be\label{pw}
\Delta(M^{1},M^{2};N^{1},N^{2})
\geq |\vec{m}_1+\vec{m}_{2}|+|\vec{m}_{1}-\vec{m}_{2}|-2.
\ee
From (\ref{pw}) one may also derive a measurement uncertainty relation which is obtained by minimizing
$\Delta(M^1,M^2,M^3;N^1,N^2,N^3)$ over pair-wise jointly measurable
measurements,
\be\label{jmerror2}
\begin{aligned}
&\Delta(M^1,M^2,M^3;N^1,N^2,N^3)\\
&=\frac{1}{2}\sum_{i<j}^{3}\Delta(M^{i},M^{j};N^{i},N^{j})\\
&\quad\geq\frac{1}{2}[\sum_{i<j}^{3}
(|\vec{m}_{i}+\vec{m}_{j}|+
|\vec{m}_{i}-\vec{m}_{j}|-2)].
\end{aligned}
\ee

Nevertheless, compared with the lower bound of (\ref{jmerror2}), the lower bound of (\ref{jmerror})
captures better incompatible measurement uncertainty of the triple measurements $M^1$, $M^2$ and $M^3$.
Consider the case that one pair of measurements $\{M^i,M^j\}$ are jointly measurable.
From the fact that
\be\label{merit}
\begin{aligned}
\sum^4_{k=1}|\vec{p}_F-\vec{p}_k|&\geq\max_{i\neq j\neq k\neq l\in \{1,2,3,4\}}(|\vec{p}_i-\vec{p}_j|+|\vec{p}_k-\vec{p}_l|)\\
&\geq2\max_{i\neq j}(|\vec{m}_i+\vec{m}_j|+|\vec{m}_i-\vec{m}_j|),
\end{aligned}
\ee
one easily gets that the lower bound of (\ref{jmerror}) is greater or equal to the lower bound of (\ref{jmerror2}).
As an example that all pairs of measurements are not jointly measurable, we consider the measurements with respect to three Pauli operators. By direct calculation we have $\mathcal{L}_1=2\sqrt{3}-2>\mathcal{L}_2=
3\sqrt{2}-3$, where $\mathcal{L}_1$ and $\mathcal{L}_2$ are the lower bounds of the inequalities (\ref{jmerror}) and (\ref{jmerror2}), respectively.
Therefore, the uncertainties from a triple of measurements are essentially different from the ones from pair wise measurements.

From (\ref{merit}) one can also analytically show that there exist triple measurements that are genuinely incompatible but pair-wise jointly measurable.
Particularly, for three measurements
$\{M^{i}_{\pm}=(I\pm\vec{m}_i\cdot\vec{\sigma})/2\}^{3}_{i=1}$,
with $\vec{m}_{1}=(1,0,0)/\sqrt{2}$, $\vec{m}_{2}=(0,1,0)/\sqrt{2}$ and $\vec{m}_{3}=(0,0,1)/\sqrt{2}$, which are proved to be pair-wise jointly measurable in \cite{Carmeli},
one verifies easily that the pair-wise lower bounds of (\ref{pw}) are all zero.
However, the lower bound of (\ref{jmerror}) is $\sqrt{6}-2>0$.

Actually, in \cite{Barchielli,Barchielli2} Barchielli et al obtained an approximation error based triple measurement uncertainty relation, where the approximation error for probabilities of joint measurements is quantified by the sum of relative entropies. Similar to $\Delta_{lb}(M^1,M^2,M^3)$, a quantity $C_{inc}(M^1,M^2,M^3)$ has been introduced in \cite{Barchielli}, although it is difficult to calculate the universal and analytical lower bound of $C_{inc}(M^1,M^2,M^3)$. In \cite{Barchielli2} a lower bound of $C_{inc}(M^1,M^2,M^3)$ has been derived for the case of three incompatible spin-$1/2$ components, which is not straightly related to the necessary and sufficient condition of the triple-wise joint measurability of the three incompatible spin-$1/2$ components.

\section{Discussion and Conclusion}
Our approach may be generalized to the case of multiple measurements. For $n$ measurements, one has
$\Delta(M^1,...,M^n;\,N^1,...,N^n)\geq \Delta_{lb}(M^1,...,M^n)$.
However, for multiple measurements the general necessary and sufficient jointly measurable conditions are still not known
even for unbiased qubit measurements.
Let us consider the multiple-wise joint measurability for arbitrary $n$ $(n\geq4)$ unbiased qubit measurements.
We have that the $n$ unbiased qubit measurements $\{N^i=(I\pm\vec{n}_i\cdot\vec{\sigma})/2\}^n_{i=1}$
are n-tuple-wise jointly measurable, if
\be\label{mjm}
\sum_{\mu_i=\pm1}|\sum^n_{i=1}\mu_i\vec{n}_i|\leq 2^n,
\ee
see proof in Appendix A.

Nevertheless, (\ref{mjm}) is not both sufficient and necessary in general. Only for some special $n$  unbiased qubit measurements $M^i$s one may have the following relation from (\ref{mjm}),
$$
\Delta(M^1,...,M^n;\,N^1,...,N^n)\geq
(\sum_{\mu_i=\pm1} |\sum^n_{i=1}\mu_i\vec{m}_i|-2^n)/2^{n-2}.
$$
Similar to the triple case, there would exist genuine incompatible $n$-tuple measurements.

By approximating a given triple of unbiased qubit measurements to all possible triple measurements that are triple-wise jointly measurable, we have formulated state-independent tight uncertainty inequalities satisfied by the triple of qubit measurements, with the lower bound giving by the necessary and sufficient condition of the triple-wise joint measurability of the given triple. These uncertainty relations can be experimentally tested, like the case of two qubit measurements \cite{Ma}. As the measurement uncertainties from a triple of measurements are essentially different from the ones from pair wise measurements, it is of significance to explore the measurement uncertainties for triple or $n$-tuple measurements by their measurement incompatibilities.
\bigskip

\noindent{\it Acknowledgements} This work is supported by the NSF of China under Grant No. 11675113 and Beijing Municipal Commission of Education (KZ201810028042). Qin acknowledges the fellowship from the China scholarship council and the support of NFSC No. 11701128. Zhang acknowledges the support of NFSC No. 11861031.

\appendix
\section{Proof of sufficient condition (\ref{mjm}) for $n$-tuple wise joint measurability}

Consider $n$ unbiased qubit measurements, $\{\frac{I+\mu_i\vec{m}_i\cdot\vec{\sigma}}{2}\}^n_{i=1}$ with $\mu_i=\pm 1$. The general measurement with measurement operators $O_{\mu_1\mu_2\cdots\mu_n}$ including  $\{\frac{I+\mu_i\vec{m}_i\cdot\vec{\sigma}}{2}\}^n_{i=1}$ as the marginal ones is given by
 \begin{widetext}
 \be
 \begin{aligned}
 &O_{\mu_1\mu_2\cdots\mu_n}\\
 &=\frac{1}{2^n}\Big[(1+\sum^{n}_{i=2}\sum_{
 \begin{aligned}
 & j_{1},j_{2},\ldots,j_{i}\in\mathcal{I}\\
 & j_{1}<j_{2}<\cdots<j_{i}
 \end{aligned}}
(\prod^{i}_{l=1}\mu_{j_l})a^{i}_{j_{1}j_{2}\cdots j_{i}})I
+(\sum^{n}_{i=2}\sum_{
 \begin{aligned}
 &j_{1},j_{2},\ldots,j_{i}\in\mathcal{I}\\
 &j_{1}<j_{2}<\cdots<j_{i}
 \end{aligned}}
(\prod^{i}_{l=1}\mu_{j_{l}})\vec{Z}^{i}_{j_{1}j_{2}\cdots j_{i}}+\sum^{n}_{i=1}\mu_{i}\vec{m}_{i})
\cdot\vec{\sigma}\Big],
\end{aligned}
\ee
\end{widetext}
where $a^i_{j_1j_2\cdots j_i}$ and $\vec{Z}^i_{j_1j_2\cdots j_i}$ are arbitrary parameters and vectors, $\forall i=1,2,\cdots,n$, $\mathcal{I}=\{1,2,\cdots,n\}$.
The positivity of the operators $\{O_{\mu_{1}\mu_{2}\cdots\mu_{n}}\}$ implies that
\be\label{ieq1}
\begin{aligned}
&\Big|\sum^{n}_{i=2}\sum_{
\begin{aligned}
&j_{1},j_{2},\ldots,j_{i}\in\mathcal{I}\\
&j_{1}<j_{2}<\cdots<j_{i}
\end{aligned}}
(\prod^{i}_{l=1}\mu_{j_{l}})\vec{Z}^{i}_{j_{1}j_{2}\cdots j_{i}}+\sum^{n}_{i=1}\mu_{i}\vec{m}_{i}\Big|\\
&\quad\leq1+\sum^{n}_{i=2}\sum_{
\begin{aligned}
&j_{1},j_{2},\ldots,j_{i}\in\mathcal{I}\\
&j_{1}<j_{2}<\cdots<j_{i}
\end{aligned}}
(\prod^{i}_{l=1}\mu_{j_{l}})a^{i}_{j_{1}j_{2}\cdots j_{i}}.
\end{aligned}
\ee

We divide the above $2^{n}$ inequalities into $2^{n-1}$ pairs such that in each pair the two inequalities take the opposite sign for all $\mu_{i}$s. From each pair of such inequalities we have the following inequality,
\be\label{ieq2}
\begin{aligned}
&\Big|\sum^{n}_{i=2,i=2t+1}
\sum_{
\begin{aligned}
&j_{1},j_{2},\ldots,j_{i}\in\mathcal{I}\\
&j_{1}<j_{2}<\cdots<j_{i}
\end{aligned}}
(\prod^{i}_{l=1}\mu_{j_{l}})\vec{Z}^{i}_{j_{1}j_{2}\cdots j_{i}}+\sum^{n}_{i=1}\mu_{i}\vec{m}_{i}\Big|\\
&\quad\leq1+\sum^{n}_{i=2,i=2t}\sum_{
\begin{aligned}
&j_{1},j_{2},\ldots,j_{i}\in\mathcal{I}\\
&j_{1}<j_{2}<\cdots<j_{i}
\end{aligned}}
(\prod^{i}_{l=1}\mu_{j_{l}})a^{i}_{j_{1}j_{2}\cdots j_{i}}.\\
\end{aligned}
\ee
Summing up all these inequalities in (\ref{ieq2}) we obtain
\begin{widetext}
\be\label{ieq3}
\sum_{\mu_{i}=\pm1}\Big|\sum^{n}_{i=2,i=2t+1}\sum_{
\begin{aligned}
&j_{1},j_{2},\ldots,j_{i}\in\mathcal{I}\\
&j_{1}<j_{2}<\cdots<j_{i}
\end{aligned}}
(\prod^{i}_{l=1}\mu_{j_{l}})\vec{Z}^{i}_{j_{1}j_{2}\cdots j_{i}}+\sum^{n}_{i=1}\mu_{i}\vec{m}_{i}\Big|
\leq 2^{n}.
\ee
Therefore, $n$ measurements $\{\frac{I\pm\vec{m}_i\cdot\vec{\sigma}}{2}\}^n_{i=1}$ are $n$-tuple wise jointly measurable if
\be\label{ieq4}
\min_{\vec{Z}^{i}_{j_{1}j_{2}\cdots j_{i}}}\sum_{\mu_{i}=\pm1}|\sum^{n}_{i=2,i=2t+1}\sum_{
\begin{aligned}
&j_{1},j_{2},\ldots,j_{i}\in\mathcal{I}\\
&j_{1}<j_{2}<\cdots<j_{i}
\end{aligned}}
(\prod^{i}_{l=1}\mu_{j_{l}})\vec{Z}^{i}_{j_{1}j_{2}\cdots j_{i}}+\sum^{n}_{i=1}\mu_{i}\vec{m}_{i}|
\leq 2^{n}.
\ee
\end{widetext}
Particularly, setting $\vec{Z}^{i}_{j_{1}j_{2}\cdots j_{i}}=0$, the above inequality reduces to
$\sum_{\mu_{i}=\pm1}|\sum^{n}_{i=1}\mu_{i}\vec{m}_{i}|\leq 2^{n}$, which assures the $n$-tuple wise jointly measurability of $\{\frac{I\pm\vec{m}_i\cdot\vec{\sigma}}{2}\}^n_{i=1}$. \qed

\end{document}